\documentclass{article}

\usepackage{PRIMEarxiv}

\usepackage[utf8]{inputenc} 
\usepackage[T1]{fontenc}    
\usepackage{hyperref}       
\usepackage{url}            
\usepackage{booktabs}       
\usepackage{amsfonts}       
\usepackage{nicefrac}       
\usepackage{microtype}      
\usepackage{lipsum}
\usepackage{fancyhdr}       
\usepackage{graphicx}       
\graphicspath{{media/}}     
\usepackage{amsmath,amssymb}
\usepackage{multirow}

\title{Treatment classification of posterior capsular opacification (PCO) using automated ground truths
}

\author{
  Raisha Shrestha\thanks{Work done while at SIIT, Thammasat University. Correspondence: \texttt{shrestha.r@unimelb.edu.au}}, Waree Kongprawechnon, Toshiaki Kondo \\
  School of Information, Computer and Communication Technology, \\ 
  Sirindhorn International Institute of Technology, \\
  Thammasat University, Pathum Thani 12000, Thailand\\
  \And
  Teesid Leelasawassuk \\
  National Electronics and Computer Technology Center (NECTEC), \\
  Pathum Thani 12120, Thailand\\
  \AND
  Nattapon Wongcumchang \\
  Department of Ophthalmology, Faculty of Medicine \\
  Thammasat University, Pathum Thani 12120, Thailand \\
  \And
  Oliver Findl, Nino Hirnschall\\
  Department of Ophthalmology, Vienna Institute for Research in Ocular Surgery, Hanusch Hospital, \\
  Vienna, Austria \\
}

\begin{document}
\maketitle

\begin{abstract}
Determination of treatment need of posterior capsular opacification (PCO)-- one of the most common complication of cataract surgery -- is a difficult process due to its local unavailability and the fact that treatment is provided only after PCO occurs in the central visual axis. In this paper we propose a deep learning (DL)-based method to first segment PCO images then classify the images into \textit{treatment required} and \textit{not yet required} cases in order to reduce frequent hospital visits. To train the model, we prepare a training image set with ground truths (GT) obtained from two strategies: (i) manual and (ii) automated. So, we have two models: (i) Model 1 (trained with image set containing manual GT) (ii) Model 2 (trained with image set containing automated GT). Both models when evaluated on validation image set gave Dice coefficient value greater than 0.8 and intersection-over-union (IoU) score greater than 0.67 in our experiments. Comparison between gold standard GT and segmented results from our models gave a Dice coefficient value greater than 0.7 and IoU score greater than 0.6 for both the models showing that automated ground truths can also result in generation of an efficient model. Comparison between our classification result and clinical classification shows 0.98 F2-score for outputs from both the models. 
\end{abstract}


\section{Introduction}
Posterior Capsular Opacification (PCO) is an after-effect of cataract surgery which causes blurriness in vision. During cataract surgery, the cloudy natural lens of a patient's eye is removed from the lens capsule and replaced with an artificial intraocular lens (IOL). Residual lens epithelial cells (LECs) in the capsular bag proliferates and move towards the posterior capsule, reaching closer to the central visual axis. This results in opacification of the posterior capsule, which makes the vision cloudy \cite{Barman2000}. The necessity of treatment is usually prescribed by any doctor after several visits of the patient only after PCO becomes prominently visible and causes visual obstruction. This study proposes an automated method to classify \textit{treatment required} and \textit{not yet required} cases of PCO which can ultimately reduce the number of follow-ups a patient has to make.

Few studies related to features, severity and grading of PCO have been carried out for analysis of PCO. EPCO \cite{Tetz1997} is one of the oldest commercial software programs which was freely available online for evaluation of PCO in terms of density and area. It was a standard technique to compare rates of PCO \cite{Aslam2002}. However, it was highly subjective and is not currently available. In \cite{Friedman1999}, an experiment was carried out to verify if retroillumination images are reliable for automated analysis by calculating density and percentage area of PCO, for which comparison with values from slit lamp images gave a good correlation. POCO system \cite{Barman2000} was texture-based and carried out objective methods to quantify PCO in terms of area but being commercially unavailable and expensive, its usability was low \cite{Aslam2002}. Likewise, AQUA \cite{Siegl2001} was an automated texture-based system which carried out fusion of images to remove illuminations. Both AQUA and POCO systems were unable to identify complex textures. Later, a PCO quantifying software called POCOman \cite{Bender2004} was built, which gave severity score and area of PCO. However, this system was highly subjective and is no longer available on the internet. A texture-based system to identify early development of PCO and determine visually significant PCO was developed by \cite{Aslam2005}. This system was updated later by OSCA \cite{Aslam2006} where light reflections were removed by combining two PCO images. \cite{Werghi2010} demonstrated a new approach of taking roughness as a texture measure for quantification of PCO. Likewise, \cite{Vivekanand2013,Vivekanand2014} considered roughness using the Hölder Exponent image to quantify PCO but it is unable to detect smooth, large pearl shaped PCO with uniform area. \cite{Grewal2008} used Pentacam tomogram images for PCO analysis based on pixel intensity and compared their results with POCOman which resulted in good correlation. Despite, this system was not practical enough, due to unavailability of scheimpflug system in maximum hospitals resulting in difficulty to receive Pentacam images. A recent system AQUA II \cite{Kronschlager2019} has successfully classified PCO into six different types using texture analysis and has automated the process of analyzing PCO. While, this system is unable to detect large single pearl shaped PCO, as it is defined by boundaries rather than textures. 

The systems developed so far have quantified PCO in terms of score, severity, area, types and intensity. Their aim is to understand PCO and its varying effects with different types of IOL so that preventive measures of PCO can be identified, lowering its occurrence. The methods used so far for PCO analysis are mostly based on conventional image processing techniques which are difficult and time consuming. Some machine learning algorithms have also been implemented,however, deep learning (DL) \cite{Lecun2015} based models have not yet been used for PCO analysis. This research for the first time proposes a DL based approach to automate the segmentation of PCO images and classify PCO in terms of treatment requirement. Despite the ongoing research on PCO, there is a lack of study which focuses on patients of PCO and their prior hospital visit before its treatment. Nd: YAG laser capsulotomy \cite{Aslam2003} which is the treatment provided for PCO, is not available in all local hospitals. Also PCO is not treated until it involves the visual axis and causes visual symptoms. As a result, patients need to visit hospital for follow ups until the treatment is provided. Therefore, to reduce these frequent follow-ups, we propose classification of PCO into \textit{treatment required} (positive PCO) and \textit{not yet required} (negative PCO) cases. In our previous research \cite{shrestha2020intensity}, we classified PCO into \textit{treatment required} and \textit{not yet required} cases by implementing a conventional method: ``intensity based method'' and an unsupervised machine learning algorithm based method: ``\textit{k}-means clustering based method''. There was a need of identification of pre-processing techniques in these methods which was difficult and critical, as it plays an important role in feature extraction of PCO. Therefore, to automate the process, we propose to implement deep learning for segmentation of PCO in this research. 

DL generally requires large dataset but for our purpose, it is difficult to gather a big dataset. Furthermore, these images require labeled ground truths which is challenging to obtain as it needs to deal with pixels rather than objects \cite{neupane2021deep}. In order to segment PCO images, we use U-Net \cite{Ronneberger2015}. It is an architecture for semantic segmentation which implements data augmentation so that limited labelled samples will be sufficient for training models eradicating the need for large training dataset. U-Net has symmetric architecture and skip connections, which allows it to give information consisting of both localization and context. This enables it to predict a reliable segmentation map.

Several medical studies have used U-Net with some variations made such as for lung segmentation \cite{Gordienko2018}, gland segmentation \cite{Chen2016}, segmentation of proximal femur \cite{Zeng2017}, etc. Because of the increasing popularity of U-Net architecture for semantic segmentation in the medical field and its advantages as mentioned above, we use U-Net for segmentation of PCO images in this research. The rest of the paper is structured as follows: Methods section describes the methods used to obtain goals of this study; Results section presents the experimental results and their validation; and Discussion section analyses the results and discusses its performance. 

\section*{Methods}\label{section2}
    \subsection*{Image Dataset}\label{section2.1}
The images used in this study were received on request from a database of PCO from Vienna, Austria. \cite{Kronschlager2019}. The same images have also been used in our previous study in \cite{shrestha2020intensity}. In total, 135 slit lamp based retro illumination images of eyes with PCO were received, out of which 118 images are selected to be used in this study. This research has considered the central 3mm area of eye as the region of interest (ROI) as it is visually significant region which determines the treatment requirement. Some of the images did not capture the central 3mm area of eyes, so this kind of images were discarded. Images with higher effect of glare were also discarded. This made a total number of 118 images for our research.

The ROI is cropped from the original images to obtain ROI-cropped images, which are split into training and test image set using \textit{k} fold cross validation technique, as we have limited number of images. Generally, the value of \textit{k} is considered as 10 or 5 because these values have shown to produce a test error rate which avoids problems of high bias and variance \cite{k-foldCross-Validation}. Hence, we chose the value of \textit{k} to be 5 allowing each fold to have approximately 23 images. In each iteration, training set has approximately 95 images and test set has approximately 23 images (4 negative case and remaining positive case images). Furthermore, this training set is split into train and validation set in a ratio of 9:1, allowing train, valid and test set to have approximately 85, 10 and 23 images respectively in each iteration.

    \subsection*{U-Net Architecture}
\begin{figure}[htp]
      \centering
      \includegraphics[width=16cm]{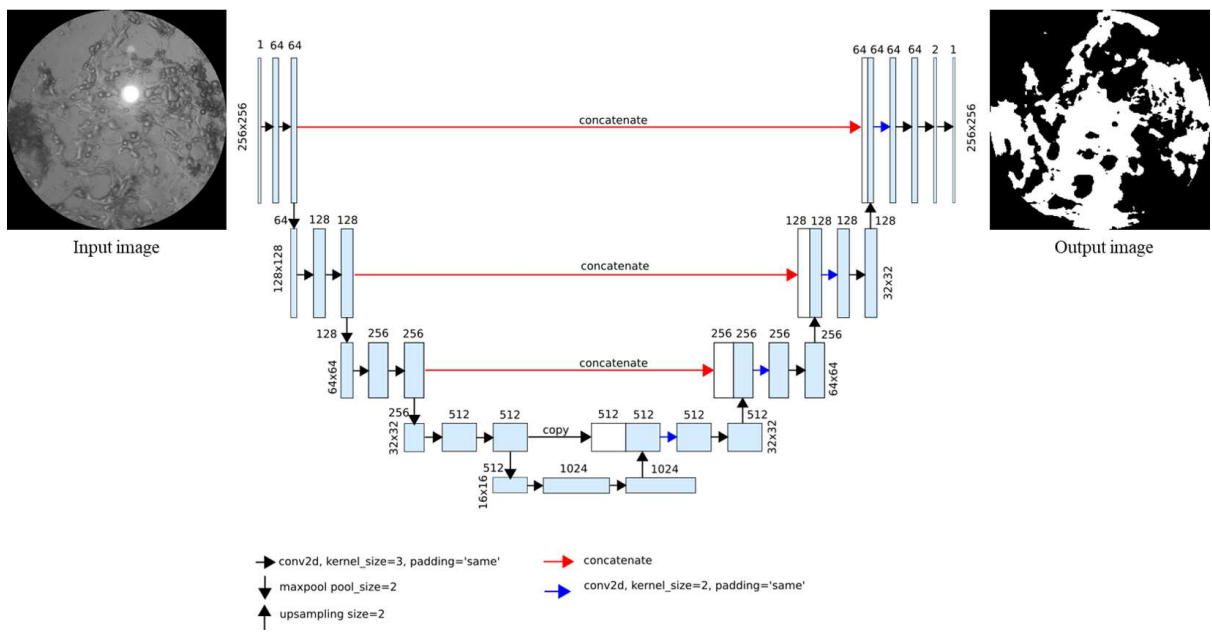}
        \caption{Network architecture of U-Net.}\label{figure1}
    \end{figure}

U-Net is a Fully Convolutional Network (FCN) generally used for semantic segmentation of biomedical images. Network architecture is U-shaped with a down-sampling path and up-sampling path as shown in Fig \ref{figure1}. Architecture of the down-sampling path follows Convolutional Neural Network (CNN). Each block consists of two 3×3 convolution layers with rectified linear unit (ReLU) as activation function and a 2×2 max pooling with stride 2. Number of feature maps is doubled at each pooling. This path collects contextual information of input image for segmentation which will be transferred to the up-sampling path by skip connections. Each block in the up-sampling path consists of a 2×2 up-convolution and two 3×3 convolutions which recovers the size of the segmentation map. Number of feature channels is halved at each up-sampling block. After each up-convolution, feature maps are concatenated which gives localization information from down-sampling path to up-sampling path. In this way, lower and higher information obtained from feature maps of each layer and blocks of the network are concatenated to produce segmentation map.

    \subsection*{Ground Truth Preparation}
Generation of a trained model using U-Net architecture require training images along with their ground truths (GTs). Different models are obtained by using two different sets of GTs which are generated manually and automatically. GT images of eyes with PCO are not available, so we generate the GT images ourselves for our entire image set. GTs are generated for our model in two ways: first is manually, under supervision of an expert ophthalmologist and second is automatically by using some image processing techniques. Manually generated GTs, which we refer as GT1 are generated using the Image Labeler tool in MATLAB where PCO regions are selected using pixel-wise labeling tools. This is a difficult and time-consuming task as PCO has complex features like shape, type and size which is why automated GTs are also generated and performance of our models trained on these two sets of GTs are evaluated.

Automated GTs which we term as GT2 are generated by using output received from a method from our previous work \cite{shrestha2020intensity} using \textit{k}-means clustering-based approach. Outputs obtained from this method have some missing PCO regions, so morphological operations like dilation and close of filter size 3×3 are applied to cover the missing regions. GT2 generation process is shown below in Fig \ref{figure2} and the images are shown in Fig \ref{figure3}.

     \begin{figure}[htp]
     \centering
     \includegraphics[width=12cm]{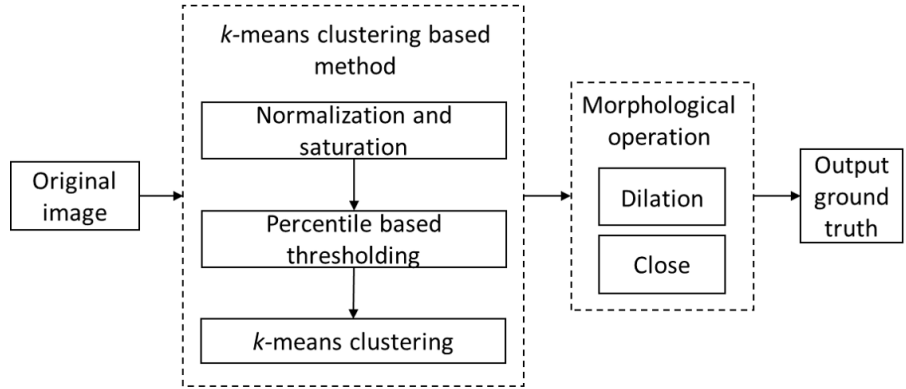}
      \caption{Automated ground truth (GT2) generation process.}\label{figure2}
     \end{figure}

     \begin{figure}[htp]
     \centering
     \includegraphics[height=12cm]{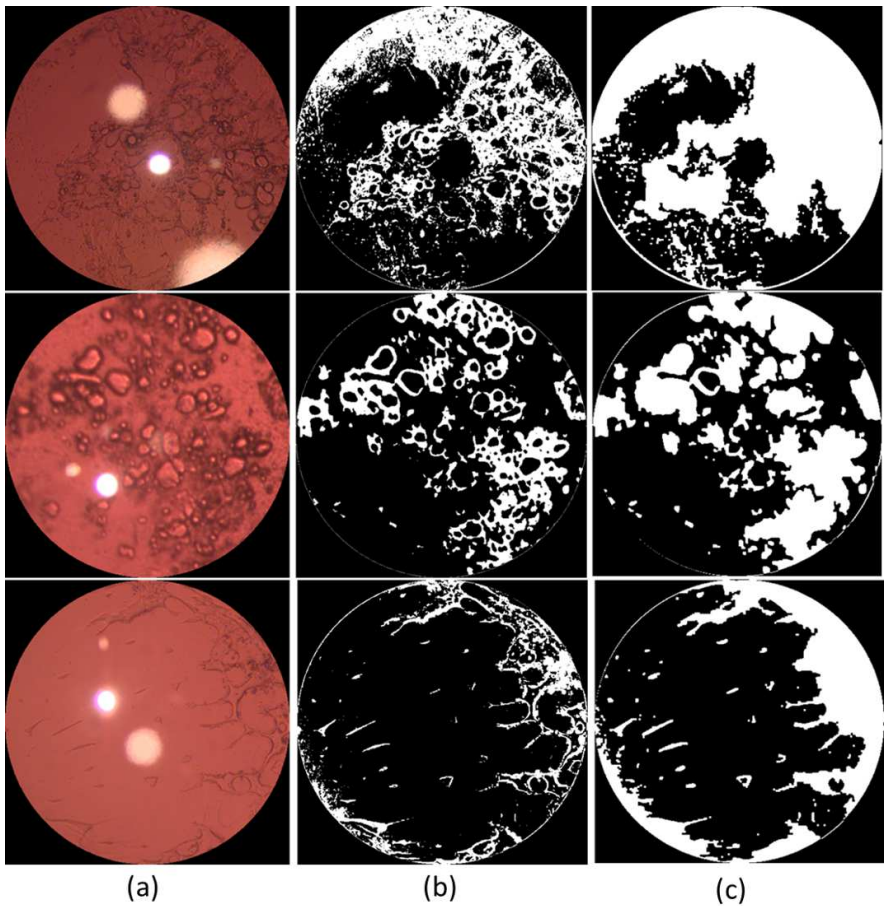}
      \caption{Automated ground truth (GT2) images. First and second rows show positive PCO cases and third row shows negative PCO case. (a) Original images. (b) Output from \textit{k}-means clustering based method \cite{shrestha2020intensity}. (c) GT2 received after application of morphological operations to Fig 3(b).}\label{figure3}
    \end{figure}

     \subsection*{Image Augmentation}\label{section2.6}
Train and valid images along with their GTs are augmented as the available image set is not enough to train a deep learning model with enough variance required by a CNN model. Operations such as rotation, width/height shift, shear, horizontal flip are applied to get augmented images. The values assigned for these operations are as follows:
    
     \begin{itemize}
    \item \textit{rotation angle:} 0.2 degrees
    \item \textit{width shift:} 0.05 fraction of total width
    \item \textit{height shift:} 0.05 fraction of total height
    \item \textit{shear range:} 0.05 degrees
         \end{itemize}

    \subsection*{Model Training and Validation}\label{section2.7}
A general network architecture of the U-Net model shown above in Fig  \ref{figure1} is used to train on our image set. Model is trained on train images along with their respective GTs. We have two sets of GTs: GT1 and GT2 as mentioned above. Model trained on train images along with GT1 is referred as Model 1 whereas model trained on train images along with GT2 is referred as Model 2. These models are validated on valid image set along with their respective GTs. Models with best performance in reference to validation Dice coefficient are saved. Overfitting is avoided by implementation of early stopping, data augmentation and \textit{k}-fold cross validation. Number of augmented images equal to the product of number of epochs, batch size and steps per epoch. These images are generated in batches during the training process. Further, saved models are used on test set to get the segmented output images.

    \subsection*{Classification of PCO Treatment Requirement}\label{section2.8}
Segmented images obtained from Model 1 and Model 2 are now evaluated for area calculation of PCO, which is measured in pixels. Among the areas of negative PCO cases belonging to GT1, a cutoff value is determined which is used to distinguish positive PCO case from negative PCO case. In order to select an optimal cutoff value from the list of negative PCO cases, we take a number of area values from negative PCO case and create confusion matrix for each of the selected values. Then, a precision-recall (PR) curve and a receiver operating characteristic (ROC) curve are drawn using these confusion matrices. With the help of these plots, we select an optimal value which gives us a higher recall as our cutoff. The output images from DL model are classified based on this cutoff value.

After talking about the method in this section, in next section we evaluate the performance of our model, as well as classification results and demonstrate the PR and ROC curves for optimal cutoff selection.

\section*{Results}\label{section3}
     \subsection*{Model Training and Performance}\label{section3.1}
Model was trained using \textit{k}-fold cross validation technique. As our chosen \textit{k} is 5, we obtained 5 models which were trained and validated on different image sets for each GT1 and GT2. We used Google Colab with Tesla T4 GPU, Intel(R) Xeon(R) CPU @ 2.20GHz and RAM 12.72 GB to train our model. Frameworks used are Keras and TensorFlow. Number of epochs, steps per epoch and batch size are chosen in such a way that training and validation loss decreases and accuracy increases. Binary cross-entropy is used as loss function and the weights are updated by Adam \cite{kingma2014adam} optimizer, with a 1e-4 learning rate. During training model, weights are saved in HDF5 format. Average time required for training each model was 1.6 hours and prediction time was 370 ms per image.

Segmented images from Model 1 is shown below in Fig \ref{figure4}. The metrics used for performance evaluation are Dice coefficient (a.k.a., F1 score), intersection-over-union (IoU) score (a.k.a., Jaccard index) and accuracy. As we used \textit{k}-fold cross validation, performance metrics for each fold were evaluated for the validation image set. The average Dice coefficient acquired is 0.81, the average IoU score is 0.68 and average accuracy is 0.87. 

Similarly, segmented images from Model 2 is shown below in Fig \ref{figure5}. Average Dice coefficient acquired is 0.8, average IoU score is 0.67 and average accuracy is 0.88. These performance metrics were obtained from evaluation on validation image set.

  \begin{figure}[htp]
      \centering
      \includegraphics[height=3.25in]{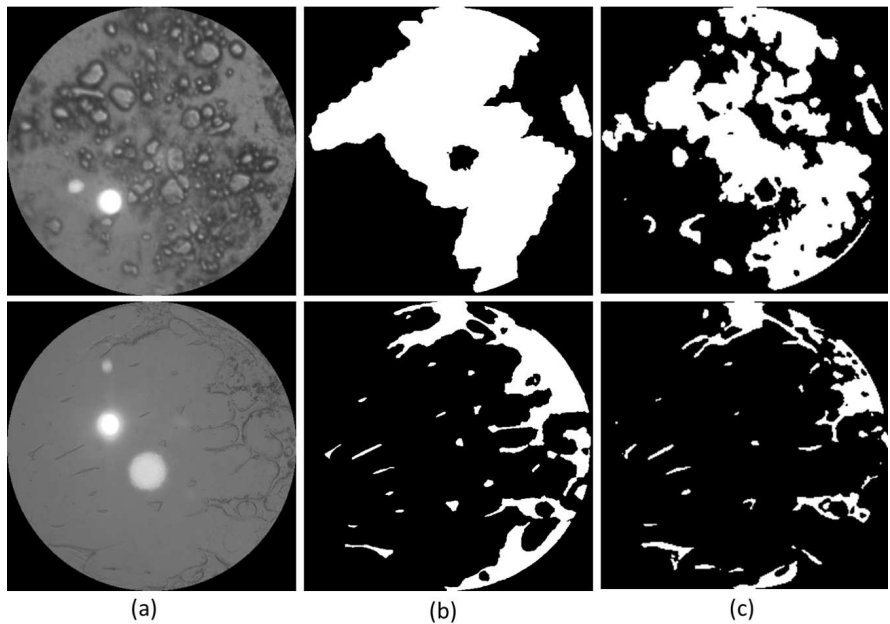}
      \caption{Segmented results from Model 1. First row shows positive   PCO and second row shows negative PCO case. (a) Input image (b) GT1, and (c) Segmented resulting images.}\label{figure4}
    \end{figure}

    \begin{figure}[htp]
      \centering
      \includegraphics[height=3.25in]{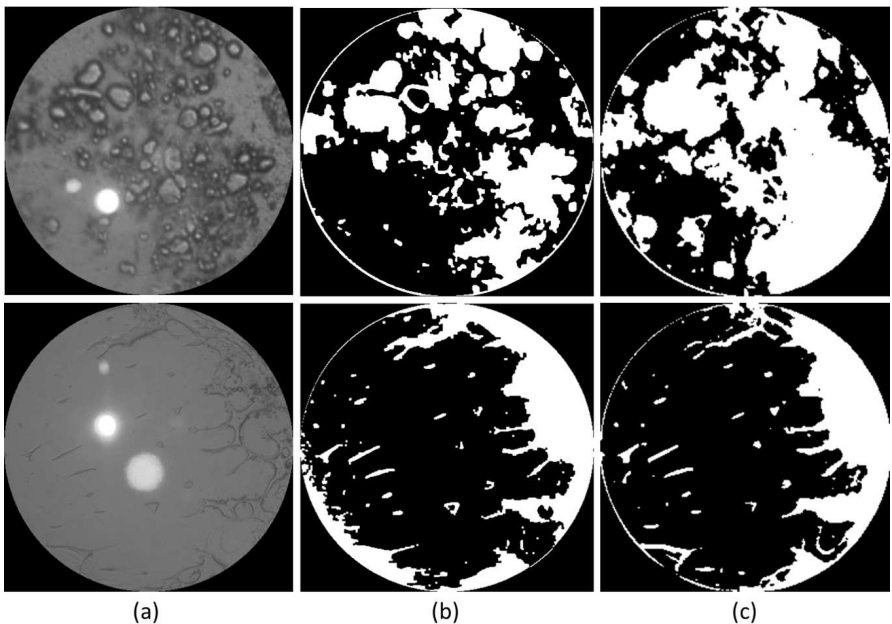}
      \caption{Segmented results from Model 2. First row shows positive PCO and second row shows negative PCO case. (a) Input image (b) GT2 (c) Segmented resulting images.}\label{figure5}
    \end{figure}

    \subsection*{Classification Results}
Area percentage of PCO in segmented images are calculated to classify PCO images into \textit{treatment required} (positive) or \textit{not yet required} (negative) PCO cases. A cutoff value is obtained from negative PCO case of GT1, which is used to classify images. Now, area of PCO for each segmented image is compared to this cutoff value. If the area of PCO for the segmented image is greater than the cutoff, then the image is classified as positive PCO case and if the area of PCO for the segmented image is smaller than the cutoff then it is classified as negative PCO case. 

\begin{figure}[ht]
    \centering
      \includegraphics[width=16cm]{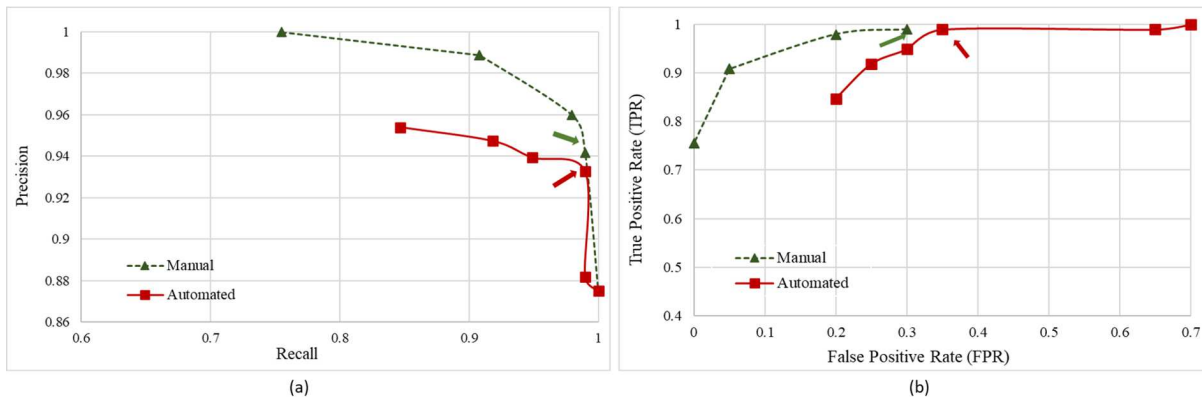}
      \caption{Curves plotting performance of models trained with GT1 and GT2 for selection of best cut-off value. (a) Precision-recall (PR) curve  (b) Receiver operator characteristics (ROC) curve.}\label{figure6}
    \end{figure}

Classification results are in binary form (0 and 1). Confusion matrices formed to draw PR and ROC curves contain metric values: True positive (TP), true negative (TN), false positive (FP) and false negative (FN). These metrics are obtained by comparing our classification with the clinical classification. The obtained confusion metrics values are used for calculation of performance metrics: recall/true positive rate (TPR), precision and false positive rate (FPR). After calculation of the performance metrics for each selected values, PR and ROC curves are plotted as shown in Fig \ref{figure6} in order to select the optimal cutoff value.

For the outputs from Model 1 in Fig 6(a), cutoff value of 4 at recall 0.989 and precision 0.94 was selected. Similarly, for outputs from Model 2, cutoff value of 12 at recall 0.95 and precision 0.94 was selected. In the ROC curve of Fig 6(b), for outputs from Model 1 and Model 2, selected cutoff value has recall/TPR 0.989 whereas for FPR, outputs from Model 1 has value 0.35 and outputs from Model 2 has value 0.3. As false negatives (FN) play a crucial role in the medical field, we chose the cutoff value which gave us higher recall/TPR such that FN is minimized.

    \subsection*{Comparison between Predicted Segmentation and GT}
\begin{figure}[htp]
      \centering
      \includegraphics[width=16cm]{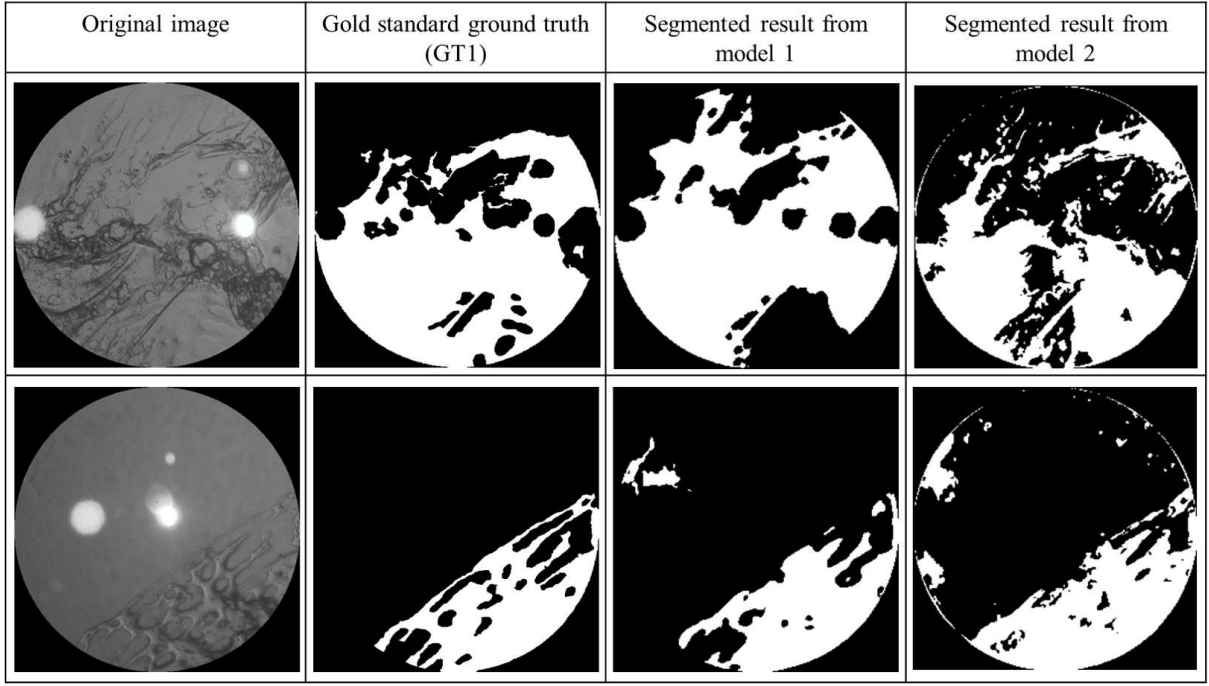}
        \caption{Comparison of gold standard ground truth, segmented result from Model 1 and segmented result from Model 2.}\label{figure7}
    \end{figure}

GT1 is the set of GT images prepared manually under expert supervision which is why it is the gold standard GT. IoU and Dice coefficient are the metrics used to evaluate the segmented images as these metrics are common and reliable in semantic segmentation to evaluate the performance of the models \cite{Metrics}. Segmented results from Model 1 when compared to the gold standard GT, gave an average Dice coefficient score of 0.74 and an average IoU score of 0.63. Likewise, comparison between segmented results from Model 2 and gold standard GT gave an average Dice coefficient value 0.73 and average IoU score 0.60. Fig \ref{figure7} shows original images of eyes with PCO, their respective gold standard GT along with segmented results from Model 1 and Model 2.

Similarly, scatterplot of area covered by PCO for predicted images obtained from Model 1 and Model 2 is shown below in Fig  \ref{figure8}. It exhibits the area values of positive and negative PCO cases, also few misclassifications (FPs and FN) are clearly spotted in the plot.

      \begin{figure}[htp]
      \centering
      \includegraphics[height=3.25in]{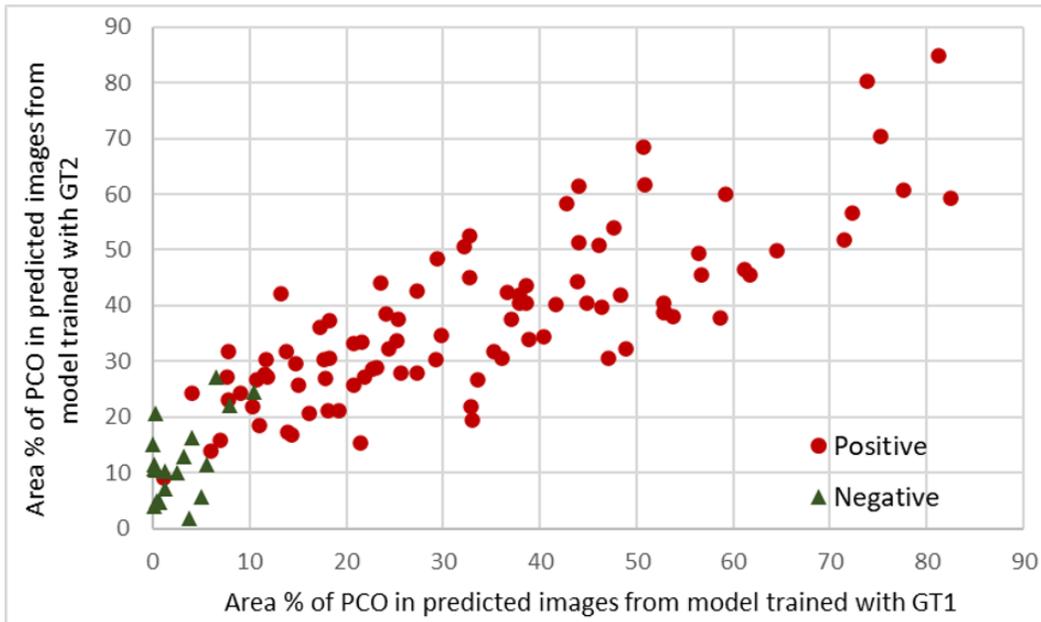}
      \caption{Scatterplot of area covered by PCO for predicted images from model trained with GT1 (Model 1) versus the predicted images from model trained with GT2 (Model 2).}\label{figure8}
    \end{figure}
    
    \subsection*{Validation}\label{section3.4}
Classification results obtained from our research is validated with classifications made by an expert ophthalmologist (clinical classification). TP, TN, FP, FN are identified and counted, then PR and ROC curves are obtained by plotting classification metrics calculated from these values. These curves help to select a cut off value which gives the best classification result. The confusion matrices for the selected cutoff values for Model 1 and Model 2 are shown in Table \ref{table1}. It shows some misclassifications through FP and FN. However, the value of FN is minimized, and the result is acceptable as FN plays a critical role in biomedical imaging.


\begin{table}[]
\centering
\caption{Confusion matrices for outputs from Model 1 and Model 2.}
\label{table1}
\begin{tabular}{|l|l|l|l|l|}
\hline
\multicolumn{1}{|c|}{\multirow{2}{*}{}} & \multicolumn{2}{c|}{Model 1} & \multicolumn{2}{c|}{Model 2} \\ \cline{2-5} 
\multicolumn{1}{|c|}{} &
  \begin{tabular}[c]{@{}l@{}}Actual \\ Positive\end{tabular} &
  \begin{tabular}[c]{@{}l@{}}Actual \\ Negative\end{tabular} &
  \begin{tabular}[c]{@{}l@{}}Actual \\ Positive\end{tabular} &
  \begin{tabular}[c]{@{}l@{}}Actual \\ Negative\end{tabular} \\ \hline
Predicted   Positive                    & 97 (TP)            & 6 (FP)           & 97 (TP)          & 7 (FP)          \\ \hline
Predicted   Negative                    & 1 (FN)             & 14 (TN)            & 1 (FN)           & 13 (TN)           \\ \hline
\end{tabular}%
\end{table}

\section*{Discussion and Conclusion}\label{section4}
In this study, PCO is segmented semantically and quantified followed by a cutoff selection which helps to classify images into \textit{treatment required} or \textit{not yet required} cases. There are a number of PCO quantifying software and studies which have been carried out as mentioned in introduction section. These research works have served for tasks like PCO analysis, quantification, severity identification and so on. However, the quantification has not been utilized to determine if the PCO needs to be treated or not. This study uses 118 PCO images collected from VEC, Austria. \cite{shrestha2020intensity} also used same images and implemented treatment classification. \cite{shrestha2020intensity} used two methods based on conventional image processing and \textit{k}-means clustering for the classification which needed significant pre-processing. Identification of a right image processing technique played an important role as it can cause loss of important features. Also, cutoff value was not selected using a standard method.
The current study considers these problems and has used deep learning-based model to segment images such that sensitive image processing task will be carried out more effectively and automatically. It also implements standard cutoff selection method by using ROC and PR curves. This research work can be a base for new research activities related to PCO which will include DL since DL has not been previously implemented for the study of PCO. 

DL based method requires a set of ground truth images which are fed to the model during the training process. We produced manual ground truths (GT1) for our entire image set. It was a time consuming as well as difficult task because of complex textures and features of PCO. So, we also aimed on generating automated GTs (GT2) and compared the performance of our models trained using both GT1 and GT2 with the gold standard GT (i.e, GT1). Comparison between the gold standard GT and the predicted results from model using GT1 (Model 1), resulted in Dice coefficient score 0.74 and IoU score 0.631. Likewise, comparing the gold standard GT with the predicted results from model using GT2 (Model 2), resulted Dice coefficient and IoU score values to be 0.73 and 0.6, respectively. Both Model 1 and Model 2 show good values of Dice coefficient and IoU scores when compared with the gold standard GT which implies that both of our models are eligible for segmentation of PCO. Performance of Model 2 also shows that automated GTs are capable of replacing the manual GTs.
    
The work of segmentation is followed by classification of PCO images into \textit{treatment required} and \textit{not yet required} cases. Cutoff values for classification is chosen with the help of PR and ROC curves as shown above in Fig \ref{figure6}. For biomedical images, false negatives play an important role. Thus, we refer to recall values and F2 scores \cite{F-score} to evaluate performance of our models. Table \ref{table2} shows the evaluation of classification through performance metrics for intensity based method \cite{shrestha2020intensity}, \textit{k}-means clustering based method \cite{shrestha2020intensity} and U-Net based models. \textit{k}-means clustering based method shows the best metric values. However, U-Net models also show promising recall and F2-score values. Intensity and \textit{k}-means clustering based methods classified PCO with performance metrics as shown in Table \ref{table2}, but these methods were semi-automated and it was challenging to extract PCO features as mentioned above. Whereas, our current approach is fully automated showing a reliable performance metric values as shown in Table \ref{table2}. Comparison between segmented results from DL models and gold standard ground truth, as well as performance metrics of classification shows that our current approach is reliable for segmentation of PCO and classification of PCO into \textit{treatment required} and \textit{not yet required} cases. 

\begin{table}[]
\centering
\caption{Evaluation of classification of PCO into \textit{treatment required} and \textit{not yet required} cases by previous methods and current study.}
\label{table2}
\resizebox{\textwidth}{!}{%
\begin{tabular}{|l|c|c|c|c|}
\hline
 &
  \begin{tabular}[c]{@{}c@{}}Intensity\\  based method \cite{shrestha2020intensity}\end{tabular} &
  \begin{tabular}[c]{@{}c@{}}k-means clustering \\ based method \cite{shrestha2020intensity} \end{tabular} &
  \begin{tabular}[c]{@{}c@{}}U-Net model \\ trained with GT1\\ (Model 1)\end{tabular} &
  \begin{tabular}[c]{@{}c@{}}U-Net model \\ trained with GT2\\ (Model 2)\end{tabular} \\ \hline
Recall    & 0.936 & 1.00  & 0.989 & 0.989 \\ \hline
Precision & 1.00  & 0.963 & 0.942 & 0.933 \\ \hline
FPR       & 0.00  & 0.214 & 0.300 & 0.350  \\ \hline
F1-score  & 0.967 & 0.981 & 0.965 & 0.960  \\ \hline
F2-score  & 0.948 & 0.99  & 0.979 & 0.978 \\ \hline
\end{tabular}%
}
\end{table}

Now, comparing outputs from our current models: Model 1 and Model 2, we can see performance metrics values of these two models do not vary much. Both the models have recall value 0.989 and F2-score greater than 0.97 showing a good classification performance. Results from segmentation and classification show that model trained with automated labels (Model 2) is equally potential of carrying out PCO segmentation and classification as the model trained with manual labels (Model 1). 
Hence the performance shown by both of our models show that DL can be implemented for automated segmentation of PCO and automated ground truths also allows the DL model to perform well for the segmentation process.

The finding of this research contributes in the study of PCO by implementing DL for PCO segmentation whereas, it can contribute in the process of providing treatment through the classification of treatment need. Implementation of this classification can ease the patients with PCO such that their number of hospital visit will be minimized. In further research, generative adversarial networks (GAN) can be used for generative modelling. GAN is capable of learning the regularities and patterns of input data and generating new data using generator and discriminator model \cite{GAN}. Implementing GAN to increase training dataset and also augmenting alongside could help the model to learn features from input data more efficiently and perform better.

    \section*{Acknowledgements}
This research was supported by the Center of Excellence in Biomedical Engineering of Thammasat University, Thailand. 

\bibliographystyle{unsrt}  
\bibliography{references}

\end{document}